\documentclass[aps,showpacs,twocolumn,floatfix]{revtex4}
\usepackage{graphicx}
\usepackage[small]{caption}
\usepackage{dcolumn}
\usepackage{longtable}
\setlongtables

\newcommand{\nc}{\newcommand}
\nc{\rnc}{\renewcommand}

\nc{\beq}{\begin{equation}}
\nc{\eeq}{\end{equation}}
\nc{\bea}{\begin{eqnarray}}
\nc{\eea}{\end{eqnarray}}
\nc{\bpi}{\begin{picture}}
\nc{\epi}{\end{picture}}
\nc{\ba}{\begin{array}}
\nc{\ea}{\end{array}}
\nc{\nn}{\nonumber}
\nc{\ts}{\textstyle}
\nc{\ds}{\displaystyle}
\nc{\bm}{\boldmath}

\nc{\p}{\partial}
\nc{\f}[2]{\frac{#1}{#2}}
\nc{\od}{{\cal O}}
\nc{\D}{{\cal D}}
\nc{\vx}{\mbox{\bm$x$}}
\nc{\vk}{\mbox{\bm$k$}}
\nc{\gh}{\hat{g}}
\nc{\pp}{\phantom{+}}
\nc{\ra}{\rightarrow}
\nc{\Fh}{\hat{F}}

\nc{\al}{\alpha}
\nc{\be}{\beta}
\nc{\ga}{\gamma}
\nc{\de}{\delta}
\nc{\ep}{\epsilon}
\nc{\ve}{\varepsilon}
\nc{\ka}{\kappa}
\nc{\om}{\omega}
\rnc{\th}{\theta}
\nc{\vp}{\varphi}
\nc{\Ga}{\Gamma}
\nc{\De}{\Delta}
\nc{\La}{\Lambda}

\nc{\eprint}[2]{#1/#2}
\nc{\gen}[3]{{\bf #1}, #2 (#3)}
\nc{\ibid}[3]{{\em ibid.} {\bf #1}, #2 (#3)}
\nc{\APNY}[3]{Ann.\ Phys.\ (N.Y.) \textbf{#1}, #2 (#3)}
\nc{\CPC}[3]{Comp.\ Phys.\ Comm.\ \textbf{#1}, #2 (#3)}
\nc{\EPJB}[3]{Eur.\ Phys.\ J.\ B \textbf{#1}, #2 (#3)}
\nc{\EPL}[3]{Europhys.\ Lett.\ \textbf{#1}, #2 (#3)}
\nc{\IJMPB}[3]{Int.\ J.\ Mod.\ Phys.\ B \textbf{#1}, #2 (#3)}
\nc{\JChP}[3]{J.\ Chem.\ Phys.\ \textbf{#1}, #2 (#3)}
\nc{\JCpP}[3]{J.\ Comput.\ Phys.\ \textbf{#1}, #2 (#3)}
\nc{\JHEP}[3]{JHEP \textbf{#1}, #2 (#3)}
\nc{\JMP}[3]{J.\ Math.\ Phys.\ \textbf{#1}, #2 (#3)}
\nc{\JPB}[3]{J.\ Phys.\ B \textbf{#1}, #2 (#3)}
\nc{\JPF}[3]{J.~Phys.~(France) \textbf{#1}, #2 (#3)}
\nc{\JTB}[3]{J.\ Theor.\ Biol.\ \textbf{#1}, #2 (#3)}
\nc{\LP}[3]{Laser Phys.\ \textbf{#1}, #2 (#3)}
\nc{\MPLB}[3]{Mod.\ Phys.\ Lett.\ B \textbf{#1}, #2 (#3)}
\nc{\MUPB}[3]{Moscow Univ.\ Phys.\ Bull.\ \textbf{#1}, #2 (#3)}
\nc{\NCD}[3]{Nuovo Cimento D \textbf{#1}, #2 (#3)}
\nc{\NCL}[3]{Nuovo Cimento Lett.\ \textbf{#1}, #2 (#3)}
\nc{\NPB}[3]{Nucl.\ Phys. B \textbf{#1}, #2 (#3)}
\nc{\PA}[3]{Physica A \textbf{#1}, #2 (#3)}
\nc{\PLA}[3]{Phys.\ Lett.\ A \textbf{#1}, #2 (#3)}
\nc{\PLB}[3]{Phys.\ Lett.\ B \textbf{#1}, #2 (#3)}
\nc{\PR}[3]{Phys.\ Rev.\ \textbf{#1}, #2 (#3)}
\nc{\PRL}[3]{Phys.\ Rev.\ Lett.\ \textbf{#1}, #2 (#3)}
\nc{\PRA}[3]{Phys.\ Rev.\ A \textbf{#1}, #2 (#3)}
\nc{\PRB}[3]{Phys.\ Rev.\ B \textbf{#1}, #2 (#3)}
\nc{\PRC}[3]{Phys.\ Rev.\ C \textbf{#1}, #2 (#3)}
\nc{\PRD}[3]{Phys.\ Rev.\ D \textbf{#1}, #2 (#3)}
\nc{\PRE}[3]{Phys.\ Rev.\ E \textbf{#1}, #2 (#3)}
\nc{\PREP}[3]{Phys.\ Rep.\ \textbf{#1}, #2 (#3)}
\nc{\RMP}[3]{Rev.\ Mod.\ Phys.\ \textbf{#1}, #2 (#3)}
\nc{\Sc}[3]{Science \textbf{#1}, #2 (#3)}
\nc{\SPJETP}[3]{Sov.\ Phys.\ JETP \textbf{#1}, #2 (#3)}
\nc{\TMF}[3]{Teor.\ Mat.\ Fiz.\ \textbf{#1}, #2 (#3)}
\nc{\ZN}[3]{Z.\ Naturforsch.\ \textbf{#1}, #2 (#3)}
\nc{\ZPB}[3]{Z.\ Phys.\ B Condens.\ Matter \textbf{#1}, #2 (#3)}

\begin{document}

\bibliographystyle{apsrev}

\title{Fluctuation pressure of a fluid membrane between walls
through six loops}

\author{Boris Kastening}
%\email[Email address: ]{kastening@oxide.tu-darmstadt.de}
\affiliation{
\mbox{Institut f\"ur Materialwissenschaft,
Technische Universit\"at Darmstadt,
Petersenstra{\ss}e 23,
D-64287 Darmstadt,
Germany}\\
and\\
\mbox{Institut f\"ur Theoretische Physik,
Freie Universit\"at Berlin,
Arnimallee 14,
D-14195 Berlin,
Germany}
email: {\tt kastening@oxide.tu-darmstadt.de}}

\date{August 2005}

\begin{abstract}
The fluctuation pressure that an infinitely extended fluid membrane
exerts on two enclosing parallel hard walls is computed.
Variational perturbation theory is used to extract the hard-wall
limit from a perturbative expansion through six loops obtained with
a smooth wall potential.
Our result $\al=0.0821\pm0.0005$ for the constant conventionally
parametrizing the pressure lies above earlier Monte Carlo results.
\end{abstract}

% 05.40.-a Fluctuation phenomena, random processes, noise,
%          and Brownian motion
% 46.70.Hg Membranes, rods and strings
% 87.16.Dg Membranes, bilayers, and vesicles
% 05.10.-a Computational methods in statistical physics
%          and nonlinear dynamics
\maketitle

\section{Introduction}

Membranes are frequent structures in chemical and biological systems.
Their dynamic behavior at finite temperature is of great interest, since
their dominant repulsive force is given by thermal out-of-plane
fluctuations \cite{HeSe,BiPeSoOs}.
If the temperature is sufficiently high, the details of the potential
that inhibits their mutual penetration or that causes them to be confined
to a certain geometrical region are unimportant.
Then the membranes' thermal fluctuations may be described by a
two-dimensional field theory with a hard-wall potential that describes
their mutual interactions and the boundary conditions of the space
accessible to them.

In an important class of membranes, their constituent molecules are able
to move freely within them.
The thermal fluctuations of these ``fluid'' membranes are controlled by
their bending rigidity $\ka$.
The curvature energy of such a membrane is, in the harmonic approximation,
described by
\beq
\label{energy}
E=\f{\ka}{2}\int_A d^2x[\p^2\vp(\vx)]^2,
\eeq
where the subscript refers to a plane with an area $A$ that serves to
parametrize the membranes' surface, and where $\vp(\vx)$ describes the
location of the membrane orthogonal to the point $\vx$ on this plane.
For the harmonic approximation to be valid, the membrane must not
fluctuate too wildly and thus the temperature must also not be too high.
It is difficult to describe the membranes' fluctuations outside
the range of validity of the harmonic approximation, since then, e.g.,
overhangs with respect to any given plane and steric self-interactions
of the membrane are possible.

There have been various theoretical approaches to compute the pressure of a
single membrane between walls \cite{HeSe,JaKl,JaKlMe,GoKr,Kl1,BaKlPe,Ka1}
or of a stack of membranes \cite{HeSe,JaKl,JaKlMe,GoKr,onlystack}.
Here we consider a fluid membrane between two rigid walls and ask what
pressure its classical statistical bending fluctuations exert on the walls.
The plane parametrizing the membrane is taken to be midway between the
enclosing walls, which are a distance $d$ apart, and we consider the limit
$A\ra\infty$.
By scaling analysis, the fluctuation pressure of the membrane has the form
\cite{HeSe}
\beq
\label{pressure}
p=\al\f{(k_BT)^2}{\ka(d/2)^3},
\eeq
and we are interested in the numerical value of $\al$.
Estimates of $\al$ range from crude theoretical estimates
$\al\approx0.0242$ by Helfrich \cite{HeSe}
and $\al\approx0.0625$ by Janke and Kleinert \cite{JaKl} (this reference
also contains an early Monte Carlo result $\al=0.060\pm0.003$) through
Monte Carlo results
\beq
\label{mc1}
\al=0.079\pm0.002
\eeq
by Janke, Kleinert, and Meinhart \cite{JaKlMe} and
\beq
\label{mc2}
\al=0.0798\pm0.0003
\eeq
by Gompper and Kroll \cite{GoKr}, and a theoretical estimate
$\al\approx0.0771$  by Kleinert \cite{Kl1} based on the analogy
with a quantum mechanical particle in a box to a theoretical
estimate $\al\approx0.0797$ by Bachmann, Kleinert, and Pelster
\cite{BaKlPe} using variational perturbation theory.
Recently, we have extended the four-loop calculation in \cite{BaKlPe}
to five loops \cite{Ka1} and found a value $\al\approx0.0820$, outside
the error bars of the Monte Carlo results.
We were, however, unable to quote an error bar for our own result.
In this work, we extend our computation through six loops.
Together with improved resummation methods, this allows us to confirm
the disagreement with the Monte Carlo results and put stringent error
bars around our result.

Our work is structured as follows.
In Sec.~\ref{bcmodel}, we briefly remind the reader how the hard
walls may be modeled using an analytic potential and how a perturbative
series for $\al$ may be derived.
In Sec.~\ref{QM}, the central results of the technically similar quantum
mechanics (QM) problem of a particle in a box are listed since they are
instrumental for extracting $\al$ for the membrane problem from its
perturbative expansion in Sec.~\ref{zeropot}.
In Sec.~\ref{oldvpt}, we directly resum the perturbative series for $\al$.
In Sec.~\ref{zeropot}, we adjust the potential modeling the boundary
conditions for the membrane problem so that the perturbative series for
$\al$ of the QM problem is obtained.
Appropriate resummation schemes let us then infer the distance of the
walls described by the resulting potential, and this information is
trivially translated into a value of $\al$ for the membrane problem.
In Sec.~\ref{summary}, we summarize and briefly discuss our results.

\section{Modeling of the hard walls}
\label{bcmodel}

Consider a tensionless membrane between two large flat parallel walls of
area $A$ separated by a distance $d$, whose curvature energy is given
by (\ref{energy}).
The $d$-dependent part $f_d$ of the free energy density of the system at
temperature $T$ is given by the functional integral
\beq
\label{fd}
\exp\left(-\f{A f_d}{k_BT}\right)
=\prod_{\vx}\int_{-d/2}^{+d/2}d\vp(\vx)\exp\left(-\f{E}{k_BT}\right).
\eeq
The pressure is then obtained as
\beq
\label{dfdd}
p=-\f{\p f_d}{\p d}
\eeq
and has the form (\ref{pressure}) \cite{HeSe,JaKl}.
Our goal is to find the numerical value of the constant $\al$.

Following an idea introduced in \cite{Kl1} and utilized also in
\cite{BaKlPe,Ka1}, we implement the restriction $-d/2<\vp<d/2$
by adding a potential term $m^4d^2\int d^2xV(\vp/d)$ to $E$,
where $V$ is an even function that is analytic inside a circle with radius
$1/2$ and has sufficiently strong singularities at $\pm1/2$.
We then expand the potential $V$ in a Taylor series in $\vp$ and drop the
restriction on $\vp$.
At the end of the calculation, we let $m\rightarrow0$ to recover the
hard-wall limit.

Since the functional form of $p$ in terms of $\ka$, $d$, and $T$ is known
and since we differentiate $f_d$ only with respect to $d$, we set
$k_BT=\ka=1$ in the sequel. 
The energy functional may then be written as
\bea
\label{epenergy}
E
&=&
\int d^2x\bigg\{\f{1}{2}[\p^2\vp(\vx)]^2+\f{1}{2}m^4\vp(\vx)^2
\nn\\
&&\qquad\quad{}
+m^4\ep_0d^2
+m^4\sum_{k=2}^\infty\ep_{2k}d^{2(1-k)}\vp(\vx)^{2k}\bigg\},\quad
\eea
where the $\ep_{2k}$ are the expansion coefficients of the potential $V$.

The above procedure defines a finite-$m$ version $f_d(m)$ of the
free energy density $f_d$ of (\ref{fd}), such that
$f_d=\lim_{m\rightarrow0}f_d(m)$.
$f_d(m)$ may be expanded in a perturbative series in terms of
vacuum diagrams---i.e., Feynman diagrams without external legs
\cite{Kl1,BaKlPe,Ka1}.
The technical details of this procedure are described in \cite{Ka1},
and deviations from the treatment in \cite{Ka1} are delegated to the
appendix.
The result is that an expansion of $f_d(m)$ through $L$ loops has the form
\beq
\label{fdm}
f_d(m)\approx\f{1}{d^2}\sum_{l=0}^L a_lg^{l-2},
\eeq
with the expansion parameter
\beq
\label{gdefmem}
g=\f{1}{m^2d^2}.
\eeq
The perturbative coefficients $a_l$ are functions of the $\ep_{2k}$.
Combining (\ref{pressure}), (\ref{dfdd}), and (\ref{fdm}), we obtain
a finite-$g$ version $\al(g)$ of $\al$ such that an expansion of $\al(g)$
through $L$ loops has the form
\beq
\label{alw}
\al(g)\approx\f{1}{4g^2}\sum_{l=0}^L a_lg^l,
\eeq
of which we need to extract the limit
\beq
\label{alpha}
\al=\lim_{g\ra\infty}\al(g).
\eeq
In Secs.~\ref{oldvpt} and \ref{zeropot}, we consider several resummation
schemes for extracting the value of $\al$ from a limited number of
coefficients $a_l$.

\section{QM Particle in a Box}
\label{QM}

A one-dimensional problem similar to the two-di\-men\-sional case above
is finding the ground state energy of a QM particle
in a one-dimensional box \cite{Kl1,KlChHa} (which, in turn, is equivalent
to finding the classical partition function of a string with tension
between one-dimensional walls \cite{Kl1,KlChHa,dGLi}).
Introduction of a potential to model the hard walls leads to a quantity
$\al(g)$ parametrizing the ground state energy of a particle moving in
this potential (see \cite{Kl1,KlChHa,Ka1} and the appendix for details;
our notation follows \cite{Ka1}).
This quantity has a loop expansion of the form (\ref{alw}) and due to
the trivial topologies of the Feynman diagrams through two loops,
the coefficients $a_0$, $a_1$, and $a_2$ are identical to those of
the membrane case.

For the particular potential
\beq
\label{vc}
V_c(z)=\f{1}{2\pi^2\cos^2(\pi z)},
\eeq
the exact ground state energy is known (see, e.g., \cite{KlChHa}) and
translates into
\beq
\label{alphac}
\al(g)=\f{\pi^2}{128}\left(\f{16}{\pi^4g^2}+\f{1}{2}
+\f{4}{\pi^2g}\sqrt{1+\f{\pi^4g^2}{64}}\right),
\eeq
giving the limit
\beq
\label{alphaqm}
\al=\f{\pi^2}{128}=0.07710628438\ldots
\eeq
for $g\ra\infty$.
The QM result (\ref{alphac}) will be utilized by the resummation schemes
of Sec.~\ref{zeropot} to extract $\al$ for the membrane problem.

In the sequel, we will always contrast the membrane results with those
for the QM problem for the same resummation scheme.

\section{\bm$\al$ from direct resummation of
\bm$\al(g)$}
\label{oldvpt}

Knowing only a few low-order coefficients $a_l$, we are looking for the
$g\ra\infty$ limit of the series (\ref{alw}).
This limit corresponds physically to removing the regulator that suppresses
fluctuations in the infrared.
In the context of critical phenomena, such series have been successfully
resummed using Kleinert's variational perturbation theory (VPT; see
\cite{Kl2,Kl3,Kl4} and Chaps.~5 and 19 of the textbooks \cite{pibook}
and \cite{phi4book}, respectively;
improving perturbation theory by a variational principle goes back at
least to \cite{Yu}).
Accurate critical exponents \cite{Kl3,Kl4,phi4book} and amplitude ratios
\cite{KlvdB} have been obtained using VPT.

In this section, we present the results of applying VPT directly to
the series (\ref{alw}) as described in \cite{BaKlPe} and in Sec.~IV
of \cite{Ka1}.
We refer the reader to \cite{Ka1} for the details and just mention
that we only present the results of the $q=1$ version of VPT, since
the results for self-consistent determination of $q$ from the series
(\ref{alw}) remain too imprecise even at the six-loop level.

\begin{figure}[t]
\begin{center}
% combinedplot.eps produced by running
% ~/membranes/math/combineplots
% in Mathematica
$~$\vspace{15pt}\\
\includegraphics[height=5cm,angle=0]{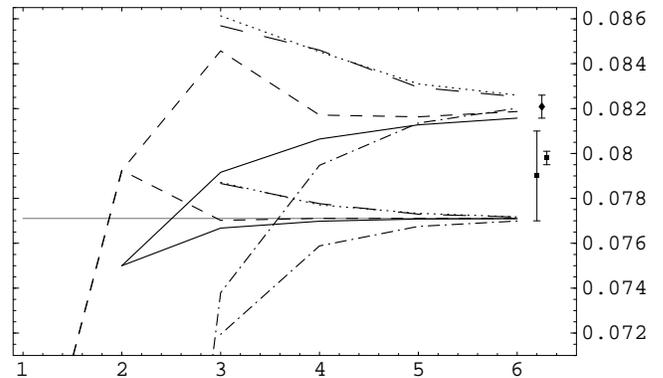}
\vspace{-10pt}
\end{center}
\caption{\protect\label{combinedplot}
$\al_{\text{qm}}$ (lower lines) and $\al_{\text{mb}}$ (upper lines) as a
function of the number of loops $L$.
The horizontal line is the exact QM result.
The dotted-dashed, short-dashed, solid, long-dashed, and dotted lines
represent data from Tables \ref{oldvptres}, \ref{alphasimple},
\ref{fixedalphaqtwo}, \ref{sclogder}, and \ref{scplat}, respectively.
Further explanations are given in the main text.
At the right, the Monte Carlo results (\ref{mc1}) and (\ref{mc2})
(boxes) as well as our final result (\ref{alpharesult}) (diamond)
are displayed.}
\end{figure}

For the potential (\ref{vc}), which also plays an important role
in the resummation variants considered in the sections below, the
$\ep_{2k}$ are listed in Table~\ref{epcoeffs}, and the corresponding
perturbative coefficients $a_l$ are listed for both the QM and the
membrane problem in Table~\ref{acoeffs}.
\begin{table}[t]
\begin{center}
\caption{\protect\label{epcoeffs}
Expansion coefficients for the potential $V_c$ and for a potential
$V_{\text{mb}}$ that gives the QM coefficients $a_l$ also for the membrane
problem.}
\begin{tabular}{l|rD{.}{.}{4.7}|D{.}{.}{4.4}}
\hline\hline
& \multicolumn{2}{c|}{$V_c$} & \multicolumn{1}{c}{$V_{\text{mb}}$} \\\hline
$\ep_0$ & $1/(2\pi^2)=$ & 0.0506606 & \multicolumn{1}{c}{same} \\
$\ep_2$ & $1/2=$ & 0.5 & \multicolumn{1}{c}{same} \\
$\ep_4$ & $\pi^2/3=$ & 3.28987 & \multicolumn{1}{c}{same} \\
$\ep_6$ & $17\pi^4/90=$ & 18.3995 & 18.0284 \\
$\ep_8$ & $31\pi^6/315=$ & 94.6129 & 89.5702 \\
$\ep_{10}$ & $691\pi^8/14175=$ & 462.545 & 419.568 \\
$\ep_{12}$ & $10922\pi^{10}/467775=$ & 2186.57 & 1890.91 \\
\hline\hline
\end{tabular}
\end{center}
\end{table}
\begin{table}[t]
\begin{center}
\caption{\protect\label{acoeffs}
Perturbative expansion coefficients for both the QM and the
membrane problem for the potential $V_c$.}
\begin{tabular}{c|rD{.}{.}{2.7}|D{.}{.}{3.10}}
\hline\hline
& \multicolumn{2}{c|}{QM}
& \multicolumn{1}{c}{membrane} \\\hline
$a_0$ & $1/2\pi^2=$      &  0.0506606 & \multicolumn{1}{c}{same} \\
$a_1$ & $1/8=$           &  0.1250000 & \multicolumn{1}{c}{same} \\
$a_2$ & $\pi^2/64=$      &  0.1542126 & \multicolumn{1}{c}{same} \\
$a_3$ & $\pi^4/1024=$    &  0.0951261 &  0.105998     \\
$a_4$ &                  &  0         &  0.026569     \\
$a_5$ & $-\pi^8/262144=$ & -0.0361959 & -0.034229     \\
$a_6$ &                  &  0         & -0.083246(13) \\
\hline\hline
\end{tabular}
\end{center}
\end{table}
The membrane's coefficients start deviating from the QM coefficients
at the three-loop level.
At the beginning, the deviation from the particular feature of the
QM series that even loop orders beyond two loops have zero coefficients
is small.
This gives the membrane's series a structure that can be expected
to be in a transitional phase towards its high-order behavior.
However, for the VPT resummation scheme to work well and give trustworthy
results, it is important that the truncated series to be resummed
resembles already the behavior at high orders.
Consequently, the dependence on the variational parameter in VPT, when
applied directly to the series (\ref{alw}), does not develop increasingly
flatter plateaus through the orders considered.
Such plateaus are, however,  an internal consistency check of the method,
and we therefore develop other resummation variants for obtaining $\al$
in the sections below.
Nevertheless, we provide in Table~\ref{oldvptres} the extension to six
loops of Eq.~(24) in \cite{Ka1} for the potential $V_c$ and for comparison
also list the corresponding QM results, taken from Table~I in \cite{Ka1}.
\begin{table}[t]
\begin{center}
\caption{\protect\label{oldvptres}
$\al$ from VPT as applied in Sec.~\ref{oldvpt} for both the QM and the
membrane problem.
$\al_{\text{mb}}$ for $L=2,3,4$ and $L=5$ were already obtained in
\cite{BaKlPe} and \cite{Ka1}, respectively.}
\begin{tabular}{D{.}{.}{2.0}|D{.}{.}{2.7}|D{.}{.}{2.7}}
\hline\hline
L & \multicolumn{1}{c|}{$\al_{\text{qm}}$}
  & \multicolumn{1}{c}{$\al_{\text{mb}}$}\\
\hline
2 & 0.0385531 & 0.0385531 \\
3 & 0.0719411 & 0.0737974 \\
4 & 0.0758821 & 0.0794726 \\
5 & 0.0767518 & 0.0813538 \\
6 & 0.0769910 & 0.0820175 \\
\hline\hline
\end{tabular}
\end{center}
\end{table}
The results are also plotted in Fig.~\ref{combinedplot} (dotted-dashed lines),
and in spite of the above critical remarks they agree perfectly well with
the results of the more refined resummation variants to be discussed below.

In \cite{YuGl}, an attempt was made to extract $\al$ from the $a_l$
through six loops using so-called factor and root approximants.
However, the achieved accuracy was not high enough for any decision about
a discrepancy with the Monte Carlo results (\ref{mc1}) and (\ref{mc2}).

\section{\bm$\al$ from zero of potential}
\label{zeropot}

Instead of resumming (\ref{alw}) directly, we apply here the strategy
of Sec.~VI of \cite{Ka1}.
That is, as a first step we fix the $\ep_{2k}$ in (\ref{epenergy}) order
by order such that the expansion of $\al(g)$ for the membrane problem
$V_{\text{mb}}$ is identical to that of the QM case with a potential $V_c$.
The resulting $\ep_{2k}$ are listed in Table \ref{epcoeffs}.
The second step is then to ask where the resulting potential
$V_{\text{mb}}(z)$ has the singularities $\pm z_0$ closest to the origin
on the real axis.
The scaling relation $f\propto1/d^2$ when $m^2=0$ allows us then
to recover $\al$ for the membrane case through
\beq
\label{almbalqm}
\al_{\text{mb}}=4z_0^2\al_{\text{qm}},
\eeq
with $\al_{\text{qm}}$ from (\ref{alphaqm}).
Since the nearest singularities of $V_c$ are of quadratic type, we 
may assume that the resulting membrane potential $V_{\text{mb}}$ has
approximately such a behavior.
We may therefore assume that $1/\sqrt{V_{\text{mb}}}$ has approximate
linear behavior at its first zero---i.e., at $\pm z_0$.

The simplest investigation of $z_0$ is to truncate the expansion
of $1/\sqrt{V}$ at $L$ loops,
\beq
\label{sqv}
1/\sqrt{V(z)}\approx\sum_{l=0}^L v_{2l}z^{2l},
\eeq
and subsequently numerically determine the first zero of the
right-hand side of (\ref{sqv}).
The $v_{2l}$ for the QM and membrane cases are listed in
Table~\ref{vcoeffs}.
\begin{table}[t]
\begin{center}
\caption{\protect\label{vcoeffs}
Expansion coefficients of (\ref{sqv}) for the quantities
$1/\sqrt{V_c}$ and $1/\sqrt{V_{\text{mb}}}$.}
\begin{tabular}{l|rD{.}{.}{3.8}|D{.}{.}{4.5}}
\hline\hline
& \multicolumn{2}{c|}{QM} & \multicolumn{1}{c}{membrane} \\ \hline
$v_0$    & $\sqrt{2}\pi=$ & 4.44288 & \multicolumn{1}{c}{same} \\
$v_2$    & $-\pi^3/\sqrt{2}=$ & -21.9247 & \multicolumn{1}{c}{same} \\
$v_4$    & $\pi^5/12\sqrt{2}=$ & 18.0324 & \multicolumn{1}{c}{same} \\
$v_6$    & $-\pi^7/360\sqrt{2}=$ & -5.93242 & 10.3394 \\
$v_8$    & $\pi^9/20160\sqrt{2}=$ & 1.04555 & -18.7293 \\
$v_{10}$ & $-\pi^{11}/1814400\sqrt{2}=$ & -0.114657 & -2.24970 \\
$v_{12}$ & $\pi^{13}/239500800\sqrt{2}=$ & 0.00857287 & 25.4600 \\
\hline\hline
\end{tabular}
\end{center}
\end{table}
The resulting values of $\al$ may be found in Table~\ref{alphasimple}.
\begin{table}[t]
\begin{center}
\caption{\protect\label{alphasimple}
Results for $\al$ for both the QM and the membrane problem using
the simple resummation scheme from the beginning of Sec.~\ref{zeropot}.}
\begin{tabular}{D{.}{.}{2.0}|D{.}{.}{2.11}|D{.}{.}{2.10}}
\hline\hline
L & \multicolumn{1}{c|}{$\al_{\text{qm}}$}
& \multicolumn{1}{c}{$\al_{\text{mb}}$} \\ \hline
1 & 0.0625000000 & \multicolumn{1}{c}{same} \\
2 & 0.0792468245 & \multicolumn{1}{c}{same} \\
3 & 0.0770188844 & 0.0845718 \\
4 & 0.0771087134 & 0.0817113 \\
5 & 0.0771062388 & 0.0816335 \\
6 & 0.0771062850 & 0.0818696(2) \\
\hline\hline
\end{tabular}
\end{center}
\end{table}
While the correct QM value (\ref{alphaqm}) is approached exponentially
fast, the convergence in the membrane case is also remarkable.
The results are plotted as short-dashed lines in Fig.~\ref{combinedplot}.

We interpret the fact that the last two differences among the
membrane values are comparable as a signal that the maximum
achievable accuracy with the current method has been reached.
A more refined approach is then needed to take into account a
likely more complicated analytic structure of the potential $V_{\text{mb}}$.
Let us therefore employ VPT to improve the naive resummation above.
Consider the quantity $\sqrt{V(z)}-\sqrt{V(0)}$.
At least in QM, the singularities of this quantity nearest to the
origin are simple poles.
The resulting series
\beq
F(z)\equiv\sqrt{V(z)}-\sqrt{V(0)}=\sum_{l=1}^\infty f_{2l}z^{2l}
\eeq
may then be inverted to
\beq
\label{z2expansion}
z^2=\sum_{l=1}^\infty u_lF^l.
\eeq
The first few coefficients $u_l$ for both the QM potential $V_c$ and
the resulting membrane potential $V_{\text{mb}}$ are listed in
Table~\ref{ucoeffs}.
\begin{table}[t]
\begin{center}
\caption{\protect\label{ucoeffs}
Expansion coefficients of (\ref{z2expansion}) for both the QM
and the membrane problem.}
\begin{tabular}{l|rD{.}{.}{4.6}|D{.}{.}{6.4}}
\hline\hline
& \multicolumn{2}{c|}{QM} & \multicolumn{1}{c}{membrane} \\ \hline
$u_1$ & $2\sqrt{2}/\pi=$ & 0.900316 & \multicolumn{1}{c}{same} \\
$u_2$ & $-10/3=$ & -3.33333 & \multicolumn{1}{c}{same} \\
$u_3$ & $128\sqrt{2}\pi/45=$ & 12.6375 & 13.1791 \\
$u_4$ & $-104\pi^2/21=$ & -48.878 & -54.6843 \\
$u_5$ & $6904\sqrt{2}\pi^3/1575=$ & 192.214 & 235.065 \\
$u_6$ & $-81784\pi^4/10395=$ & -766.379 & -1037.10 \\
\hline\hline
\end{tabular}
\end{center}
\end{table}
We are interested in finding
\beq
z_0^2=\lim_{F\ra\infty}z^2(F).
\eeq

Motivated by the successes of such an ansatz in critical phenomena,
we assume that the function $F$ can be expanded around its first
singularities $\pm z_0$ as
\beq
\label{fexp}
F=\sum_{k=0}^\infty\bar{u}_k(z^2-z_0^2)^{-q/2+k},
\eeq
where $q=2$ for QM.
Inversion of (\ref{fexp}) gives
\beq
z^2=\sum_{m=0}^\infty u'_mF^{-2m/q},
\eeq
with $z_0^2=u'_0$.
We may either set $q=2$ as in QM in the hope that the deviation for the
membrane case is small, or determine $q$ self-consistently.
We use both approaches below.

Now apply VPT \cite{pibook,phi4book}.
In a truncated expansion
\beq
z^2\approx\sum_{l=1}^Lu_lF^l,
\eeq
we replace
\bea
F^l
&\ra&
(tF)^l\left\{\left(\f{F}{\Fh}\right)^{2/q}
+t\left[1-\left(\f{F}{\Fh}\right)^{2/q}\right]\right\}^{-lq/2}
\nn\\
&=&
(t\Fh)^l
\left\{1+t\left[\left(\f{\Fh}{F}\right)^{2/q}-1\right]\right\}^{-lq/2},
\eea
reexpand the resulting expression in $t$ through $t^L$, set $t=1$,
and then optimize the resulting expression in $\Fh$, where optimizing
refers to finding appropriate stationary or turning points according to
the principle of minimal sensitivity \cite{St}.
That is, we replace
\beq
\label{vvhatIII}
F^l
\ra
\Fh^l\sum_{k=0}^{L-l}
\left(\ba{c}-lq/2\\k\ea\right)
\left[\left(\f{\Fh}{F}\right)^{2/q}-1\right]^k
\eeq
and optimize the resulting expression in $\Fh$.
In the limit $F\ra\infty$ of interest to us, this amounts to
\beq
\label{z02}
z_0^2\approx\text{opt}_{\Fh}\left[\sum_{l=1}^Lu_l\Fh^l\sum_{k=0}^{L-l}
\left(\ba{c}-lq/2\\k\ea\right)(-1)^k\right],
\eeq
which is the $L$-loop approximation to $z_0^2$---i.e., using the
expansion coefficients through $u_L$.
It turns out that, through the order we are working, there is exactly
one extremum for even $L$ and exactly one turning point and no extremum
for odd $L$.
This makes the choice of the optimization unique at each order.
The value of $\al$ is in each case obtained through (\ref{almbalqm}).

The results for $q=2$ are summarized in Table~\ref{fixedalphaqtwo}
and plotted as solid lines in Fig.~\ref{combinedplot}.
\begin{table}[t]
\begin{center}
\caption{\protect\label{fixedalphaqtwo}
Results for $\al$ when the $a_l$ are fixed to be those of the QM problem
and $q=2$ is assumed.}
\begin{tabular}{D{.}{.}{2.0}|D{.}{.}{2.8}|D{.}{.}{2.8}}
\hline\hline
L & \multicolumn{1}{c|}{$\al_{\text{qm}}$}
& \multicolumn{1}{c}{$\al_{\text{mb}}$} \\\hline
2 & 0.0750000 & 0.0750000 \\
3 & 0.0766754 & 0.0791616 \\
4 & 0.0769828 & 0.0806435 \\
5 & 0.0770794 & 0.0812768 \\
6 & 0.0770973 & 0.0815743 \\
\hline\hline
\end{tabular}
\end{center}
\end{table}
The correct QM value (\ref{alphaqm}) is approached exponentially fast.
The convergence in the membrane case is also remarkable.
Though the values are slightly lower than those reported in
Tables~\ref{oldvptres} and \ref{alphasimple}, they clearly point towards
a value of $\al$ above the results (\ref{mc1}) and (\ref{mc2}).

If we refrain from making assumptions about $q$ for $z^2(F)$, we can
determine it self-consistently by treating first $d\ln z^2/d\ln F$ in VPT
\cite{Kl3,phi4book}, since it has the same $q$ as $z^2(F)$ and since
\beq
\label{logder}
\lim_{F\ra\infty}\f{d\ln z^2}{d\ln F}=0
\eeq
by the assumption of a singularity of the potential.
That is, we resum the expansion of $d\ln z^2/d\ln F$ as detailed above
and tune $q$ such that optimization with respect to $\Fh$ leads to
(\ref{logder}).
Through two loops, the expansion of $d\ln z^2/d\ln F$ is $q$-independent,
and we start with $L=3$.
It turns out that through the order we are working, we must use
turning points for even $L$ and maxima for odd $L$ when determining $q$.
For subsequently determining $z_0^2$, the situation is reverse---namely,
as above for $q=2$.

The results for $q$ and $\al$ through six loops are listed in
Table~\ref{sclogder}.
The results for $\al$ are plotted as long-dashed lines in
Fig.~\ref{combinedplot}.
\begin{table}[t]
\begin{center}
\caption{\protect\label{sclogder}
Results for $q$ and $\al$ when the $a_l$ are fixed to be those of
the QM problem and $q$ is determined from its own resummed series.}
\begin{tabular}{D{.}{.}{2.1}|D{.}{.}{2.5}D{.}{.}{2.7}
|D{.}{.}{2.8}D{.}{.}{2.10}}
\hline\hline
L & \multicolumn{1}{c}{$q_{\text{qm}}$}
  & \multicolumn{1}{c|}{$\al_{\text{qm}}$}
  & \multicolumn{1}{c}{$q_{\text{mb}}$}
  & \multicolumn{1}{c}{$\al_{\text{mb}}$}
\\\hline
3 & 2.09487 & 0.0786643 & 2.26290    & 0.0856888    \\
4 & 2.05356 & 0.0777648 & 2.21951    & 0.0846057    \\
5 & 2.02049 & 0.0772965 & 2.11817    & 0.0829441    \\
6 & 2.00822 & 0.0771659 & 2.08666(1) & 0.0825299(2) \\
\hline\hline
\end{tabular}
\end{center}
\end{table}
Note how $q$ approaches $2$ rapidly for the QM problem and that also
for the membrane problem a value around $2$ appears to be approached.

An alternative to using (\ref{logder}) for the determination of $q$ is
to tune $q$ such that the plateaus at which the result depends least on
variations of $\Fh$ are optimized \cite{HaKl}.
This strategy has been successfully applied in \cite{HaKl,Ka2} in the
context of critical phenomena.
In practice, this means finding $\Fh$ and $q$ such that first and second
derivatives of the right-hand side of (\ref{z02}) with respect to $\Fh$
vanish (for turning points) or such that first and third derivatives with
respect to $\Fh$ vanish (for extrema).

The results for $q$ and $\al$ through six loops are listed in
Table~\ref{scplat} and are very similar to those of Table~\ref{sclogder}.
\begin{table}[t]
\begin{center}
\caption{\protect\label{scplat}
Results for $q$ and $\al$ when the $a_l$ are fixed to be those of
the QM problem and $q$ is determined from optimized plateaus.}
\begin{tabular}{D{.}{.}{2.1}|D{.}{.}{2.5}D{.}{.}{2.7}
|D{.}{.}{2.8}D{.}{.}{2.10}}
\hline\hline
L & \multicolumn{1}{c}{$q_{\text{qm}}$}
  & \multicolumn{1}{c|}{$\al_{\text{qm}}$}
  & \multicolumn{1}{c}{$q_{\text{mb}}$}
  & \multicolumn{1}{c}{$\al_{\text{mb}}$}
\\\hline
3 & 2.09730 & 0.0787132 & 2.28225    & 0.0861317    \\
4 & 2.04990 & 0.0777101 & 2.21532    & 0.0845332    \\
5 & 2.02405 & 0.0773337 & 2.13061    & 0.0830987    \\
6 & 2.00948 & 0.0771767 & 2.09303(1) & 0.0825984(2) \\
\hline\hline
\end{tabular}
\end{center}
\end{table}
They are plotted as dotted lines in Fig.~\ref{combinedplot}.

\section{Summary and Discussion}
\label{summary}

In summary, we have computed the constant $\al$ parametrizing the
pressure law (\ref{pressure}) of an infinitely extended fluid membrane
between two parallel hard walls.
The hard wall was replaced by a smooth potential, allowing for a
perturbative loop expansion for $\al$.
Several resummation schemes were used to extract the hard-wall limit
from expansion coefficients through six loops with results listed in
Tables \ref{oldvptres}, \ref{alphasimple}, \ref{fixedalphaqtwo},
\ref{sclogder}, and \ref{scplat} and plotted in Fig.~\ref{combinedplot}.

The values from Table~\ref{fixedalphaqtwo} on the one hand and
Tables~\ref{sclogder} and \ref{scplat} on the other hand approach
each other with increasing numbers of loops from below and above,
respectively.
A conservative procedure for combining our results for $\al$ is
to average the lowest and highest six-loop values for $\al$ obtained
above (i.e., the values for $\al$ from
Tables~\ref{fixedalphaqtwo} and \ref{scplat}, respectively) and take
their difference to be the full error bar.
This provides our final result
\beq
\label{alpharesult}
\al=0.0821\pm0.0005,
\eeq
displayed in Fig.~\ref{combinedplot}.
It lies above the Monte Carlo results (\ref{mc1}) and (\ref{mc2}),
also displayed in Fig.~\ref{combinedplot}.
The simplest explanation we have to offer for this discrepancy is that
their error bars, in particular that of (\ref{mc2}), may have been chosen
too optimistic.
Also, finite-size or other systematic effects may not have been taken into
account properly.

On the other hand, it is possible that the treatment in our work
is inflicted by systematic errors.
Experience with VPT tells that the internal consistency checks,
especially the development of increasingly flatter plateaus in the
optimization procedure with higher orders, are reliable indicators of
VPT to work.
These checks have successfully been implemented for the procedures of
Sec.~\ref{zeropot}.
Nevertheless, it cannot be ruled out that the method used here is not
flexible enough to adequately take into account the unknown true
analytical structure of $\al(g)$.

Another concern is the value of $q_{\text{mb}}$.
While the treatments leading to the $\al$ values listed in
Tables \ref{sclogder} and \ref{scplat} and thus leading to the upper
dotted and long-dashed curves in Fig.~\ref{combinedplot} determine
$q_{\text{mb}}$ self-consistently, the QM-inspired value $q_{\text{mb}}=2$
was somewhat arbitrarily chosen to obtain the values for $\al$ in
Table~\ref{fixedalphaqtwo}.
What if the value of $q_{\text{mb}}$ describing $\al(g)$ best differs
from 2?
A slightly larger value (but below the ones from the self-consistent
determinations of $q_{\text{mb}}$) leads to slightly increased values
of $\al$ and therefore also to a larger mean value and smaller error bar
in (\ref{alpharesult}).
On the other hand, a slightly smaller value leads to slightly decreased
values of $\al$ and therefore also to a smaller mean value and larger
error bar in (\ref{alpharesult}).
In both cases, the $q_{\text{mb}}=2$ curve may still approach the correct
result.
For an optimal value of $q_{\text{mb}}$ below 2 this may happen by
decreasing $\al$ values at higher orders (that such a behavior is possible
in principle is easily tested by setting $q_{\text{mb}}$ to a value above
$2.085$, which causes the corresponding six-loop value for $\al$ to drop
below that at five loops).
The smooth behavior and slow flattening of the $q_{\text{mb}}=2$
curve lets us believe, though, that such a drop, if present, should be
very small.
The best $q_{\text{mb}}$ value can thus be expected to lie at
most slightly below 2, leading to only a small decrease of the values
of Table~\ref{fixedalphaqtwo} and the upper solid curve in
Fig.~\ref{combinedplot} and to a slightly lower mean value and larger error
bar than given in (\ref{alpharesult}).
It is reassuring that the results of both the direct resummation
of $\al(g)$ employed in Sec.~\ref{oldvpt} and the naive resummation
from the beginning of Sec.~\ref{zeropot} lie very close to the mean
value of (\ref{alpharesult}).

All things considered, we are rather confident about our result
(\ref{alpharesult}), but further studies of the system are required
to settle the question of the correct value of $\al$.

\appendix

\section*{Appendix: Feynman diagrams}
\label{loopexp}

In \cite{Ka1}, we have described at length how recursion relations along
the lines of \cite{recrel1,recrel2} can be used to construct the vacuum
diagrams needed for the computation of the perturbative coefficients
$a_l$ in (\ref{alw}) through a given loop order, and there is no need to
repeat the derivation here.
Given the Feynman diagrams, a vertex with $2k$ lines represents a factor
\beq
\label{ccfactor}
-m^4d^{2(1-k)}\ep_{2k},
\eeq
and the perturbative coefficients $a_l$ are obtained from the sum of
$l$-loop diagrams as
\beq
a_l=-\sum_nc_{l\text{-}n}g_{l\text{-}n}I_{l\text{-}n},
\eeq
where $c_{l\text{-}n}$ is a combinatorial factor, $g_{l\text{-}n}$ is a
monomial in the $\ep_{2k}$, and $I_{l\text{-}n}$ is the corresponding
momentum space integral.
The integration measure is $\int d^Dk/(2\pi)^D$ with $D=1$ for QM and
$D=2$ for the membrane.
The membrane propagator carrying a momentum $\vk$ is given by
$1/(k^4+m^4)$, while the QM propagator carrying a momentum $k$ is
given by $1/(k^2+m^2)$.

From the list of diagrams through five loops in \cite{Ka1} it is obvious
that a major reason for the rapid increase of the number of Feynman
diagrams with the number of loops is a proliferation of the
momentum-independent one-loop propagator insertions of the form
\bpi(18,0)
\put(5,-2.5){\line(1,0){8}}
\put(9,3){\circle{11}}
\put(9,-2.5){\circle*{2}}
\epi.
A simple measure to reduce the number of diagrams to be considered is
a one-loop resummation where we absorb the above insertion into the
parameter $m$ in the propagator and at the end reexpand the resulting
modified perturbative series in powers of $g$ as was carried out in
\cite{Ka3} and treated on a more formal level in \cite{recrel2}.
Note that many other resummations in the form of momentum-independent
propagator and vertex corrections are possible, leading to a further
reduction of the number of Feynman diagrams.
However, at the current level of computing vacuum diagrams through six
loops, this is unnecessary.
In any case, the diagrams most difficult to evaluate are always those with
the full loop topology and remain after any such resummation.

To implement the one-loop resummation in the membrane case, we must
compute diagrams with a modified propagator $H$ such that
\beq
\label{olrsdef}
G^{-1}_{12}=H^{-1}_{12}-12L^{(4)}_{1234}H_{34},
\eeq
where the notation of \cite{Ka1} has been employed.
Writing $G^{-1}=k^4+m^4$ and $H^{-1}=k^4+M^4$, this condition translates
into
\beq\ts
k^4+m^4=k^4+M^4-\f{3}{2}m^4d^{-2}\ep_4M^{-2}
\eeq
or
\beq
\label{Mmmemb}
\left(\f{M^2}{m^2}\right)^3-\left(\f{M^2}{m^2}\right)-\f{3}{2}\ep_4g=0,
\eeq
where we have used (\ref{gdefmem}) and the one-loop result
\beq
\int\f{d^2k}{(2\pi)^2}\f{1}{k^4+m^4}=\f{1}{8m^2}.
\eeq
Define $Z(g)$ by
\beq
M^2=Z(g)m^2.
\eeq
Although (\ref{Mmmemb}) can be solved analytically for $Z(g)$, it is
more useful for our purposes to write
\beq
Z(g)=1+\sum_{k=1}^\infty c_kg^k
\eeq
and extract
\beq\ts
c_1=\f{3}{4}\ep_4
\eeq
and the recursion relation
\beq
c_k=-\f{3}{2}\sum_{i=1}^{k-1}c_ic_{k-i}
-\f{1}{2}\sum_{i=1}^{k-2}\sum_{j=1}^{k-i}c_{k-i-j}c_ic_j,\quad k>1.
\eeq
Through six loops in the vacuum diagrams, we need $Z(g)$ through $g^5$
and obtain
\bea
Z(g)
&=&\ts
1+\f{3}{4}\ep_4g-\f{27}{32}\ep_4^2g^2+\f{27}{16}\ep_4^3g^3
-\f{8505}{2048}\ep_4^4g^4
\nn\\
&&\ts{}
+\f{729}{64}\ep_4^5g^5+\od(g^6).
\eea

The same resummation (\ref{olrsdef}) can be implemented for the QM case.
Writing $G^{-1}=k^2+m^2$ and $H^{-1}=k^2+M^2$, the condition
(\ref{olrsdef}) translates now into
\bea
k^2+m^2=k^2+M^2-6m^2d^{-2}\ep_4M^{-1}
\eea
or
\beq
\label{Mmqm}
\left(\f{M}{m}\right)^3-\left(\f{M}{m}\right)-\f{3}{2}\ep_4g=0,
\eeq
where we have used $g=4/md^2$ \cite{Ka1} and the one-loop result
\beq
\int_{-\infty}^{+\infty}\f{dk}{2\pi}\f{1}{k^2+m^2}=\f{1}{2m}.
\eeq
This time we define $Z(g)\equiv M/m$.
Then $Z(g)$ is the same as in the membrane case considered above.

The resulting modifications in the Feynman rules are for both the
membrane and the QM case:
\begin{enumerate}
\item
Discard all vacuum diagrams with one or more one-loop propagator
insertions
\bpi(18,0)
\put(5,-2.5){\line(1,0){8}}
\put(9,3){\circle{11}}
\put(9,-2.5){\circle*{2}}
\epi,
with the exception of the two-loop diagram, which changes sign (here,
the combinatorics do not work out; loosely speaking, it is undefined which
part of the diagram is the insertion).
\item
Compute all remaining diagrams with the replacement $m\ra M$
in the propagators.
\item
Replace $M^2\ra Z(g)m^2$ for the membrane case and $M\ra Z(g)m$ for the
QM case and reexpand the perturbative series in powers of $g$.
\end{enumerate}

In Table~\ref{numbersofdiagrams}, we give the original numbers of
diagrams at some low loop orders, the numbers left after our one-loop
resummation, and the numbers of diagrams with the full respective
loop topology.
The latter is necessarily the same both before and after the one-loop
resummation.
\begin{table}[t]
\begin{center}
\caption{\protect\label{numbersofdiagrams}
Numbers of diagrams for low loop orders.}
\begin{tabular}{c|D{.}{.}{2.0}D{.}{.}{1.0}D{.}{.}{1.0}D{.}{.}{1.0}
D{.}{.}{1.0}D{.}{.}{2.0}D{.}{.}{2.0}D{.}{.}{3.0}}
\hline\hline
number of loops $l$
& 0 & 1 & 2 & 3 & 4 & 5 & 6 & 7\\\hline
diagrams
& 1 & 1 & 1 & 3 & 7 & 24 & 83 & 376\\
diags.\ after one-loop resum.\
& 1 & 1 & 1 & 2 & 3 & 11 & 29 & 125\\
diags.\ with $l$-loop topology & 1 & 1 & 0 & 1 & 1 & 5 & 8 & 37 \\
\hline\hline
\end{tabular}
\end{center}
\end{table}
In Table~\ref{diagrams}, we list all diagrams through six loops left
after the one-loop resummation.
Also given are their combinatorial factors $c_{l\text{-}n}$, their coupling
constant factors $g_{l\text{-}n}$, and the values $I_{l\text{-}n}$ of the
corresponding integrals for $M=1$ for both the QM and the membrane problem
(the multiplying power of $M$ can immediately be inferred from the number
of loops and propagators of a given diagram).

The techniques for evaluating the integrals are explained in \cite{Ka1}.
With the exception of $I_{6\text{-}5}$, the membrane integrals have been
evaluated to the precision given either in momentum space or in both
momentum and configuration space.
For $I_{6\text{-}5}$, the indicated precision could only be obtained in
configuration space.
Since the slightly lower precision of $I_{6\text{-}5}$ introduces the main
computational error into the determination of $\al$ at the six-loop
level, we have indicated the ensuing numerical error in the other
tables of this work, where applicable.
%
%\newpage
\begingroup
\squeezetable
\begin{longtable}[c]{c|c|c|c|c|c}
\caption{\label{diagrams}
Diagrams $l\text{-}n$ ($n$th $l$-loop diagram) through six loops,
their combinatorial factors $c_{l\text{-}n}$, coupling constant factors
$g_{l\text{-}n}$, and values $I_{l\text{-}n}$ of the corresponding
integrals for $M=1$.
$D=1$ and $D=2$ correspond to the QM and membrane problems, respectively.}
\\
$l{-}n$&diagram&$c_{l\text{-}n}$&$g_{l\text{-}n}$&$I_{l\text{-}n}^{D=1}$
&$I_{l\text{-}n}^{D=2}$
\\\hline\hline\hline
\endfirsthead
$l{-}n$&diagram&$c_{l\text{-}n}$&$g_{l\text{-}n}$&$I_{l\text{-}n}^{D=1}$
&$I_{l\text{-}n}^{D=2}$
\\\hline\hline\hline
\endhead
0-1 &
\rule[-2pt]{0pt}{10pt}
\bpi(10,0)
\put(5,3){\circle*{2}}
\epi
& $1$ & $-\ep_0$ & $1$ & $1$
\\\hline\hline
1-1 &
\rule[-10pt]{0pt}{26pt}
\bpi(26,0)
\put(13,3){\circle{16}}
\epi
& $\ds\f{1}{2}$ & $1$ & $-1$ & $\ds-\f{1}{4}$
\\\hline\hline
2-1 &
\rule[-10pt]{0pt}{26pt}
\bpi(42,0)
\put(13,3){\circle{16}}
\put(29,3){\circle{16}}
\put(21,3){\circle*{2}}
\epi
& $-3$ & $-\ep_4$ & $\ds\f{1}{4}$ & $\ds\f{1}{64}$
\\\hline\hline
3-1 &
\rule[-14pt]{0pt}{34pt}
\bpi(34,0)
\put(17,3){\circle{24}}
\put(17,3){\oval(24,8)}
\put(5,3){\circle*{2}}
\put(29,3){\circle*{2}}
\epi
&$12$&$\ep_4^2$&$\ds\f{1}{32}$&$4.04576\times10^{-4}$
\\\hline
3-2 &
\rule[-13pt]{0pt}{37pt}
\bpi(42,0)
\put(21,3){\circle*{2}}
\qbezier(21,3)(13.79,10.21)(13.79,13.79)
\qbezier(13.79,13.79)(13.79,21.)(21.,21.)
\qbezier(21,3)(28.21,10.21)(28.21,13.79)
\qbezier(28.21,13.79)(28.21,21.)(21.,21.)
\qbezier(21,3)(18.36,-6.85)(15.26,-8.64)
\qbezier(15.26,-8.64)(9.02,-12.24)(5.41,-6.)
\qbezier(21,3)(11.15,5.64)(8.05,3.85)
\qbezier(8.05,3.85)(1.81,0.24)(5.41,-6.)
\qbezier(21,3)(30.85,5.64)(33.95,3.85)
\qbezier(33.95,3.85)(40.19,0.24)(36.59,-6.)
\qbezier(21,3)(23.64,-6.85)(26.74,-8.64)
\qbezier(26.74,-8.64)(32.98,-12.24)(36.59,-6.)
\epi
&$ 15 $&$ -\ep_6 $&$\ds\f{1}{8}$&$\ds\f{1}{512}$
\\\hline\hline
4-1 &
\rule[-14pt]{0pt}{34pt}
\bpi(34,0)
\put(17,3){\circle{24}}
\put(6.6,9){\line(1,0){20.8}}
\put(6.6,9){\line(3,-5){10.4}}
\put(27.4,9){\line(-3,-5){10.4}}
\put(6.6,9){\circle*{2}}
\put(27.4,9){\circle*{2}}
\put(17,-9){\circle*{2}}
\epi
&$ 288 $&$ -\ep_4^3 $&$\ds\f{3}{512}$&$ 1.63237\times10^{-5} $
\\\hline
4-2 &
\rule[-14pt]{0pt}{34pt}
\bpi(50,0)
\put(17,3){\circle{24}}
\put(17,3){\oval(24,8)}
\put(5,3){\circle*{2}}
\put(29,3){\circle*{2}}
\put(37,3){\circle{16}}
\epi
&$ 360 $&$ \ep_4\ep_6 $&$\ds\f{1}{64}$&$ 5.05719\times10^{-5} $
\\\hline
4-3 &
\rule[-16pt]{0pt}{38pt}
\bpi(42,0)
\put(21,3){\circle*{2}}
\qbezier(21,3)(21.,13.2)(23.53,15.73)
\qbezier(23.53,15.73)(28.63,20.83)(33.73,15.73)
\qbezier(21,3)(31.2,3.)(33.73,5.53)
\qbezier(33.73,5.53)(38.83,10.63)(33.73,15.73)
\qbezier(21,3)(10.8,3.)(8.27,5.53)
\qbezier(8.27,5.53)(3.17,10.63)(8.27,15.73)
\qbezier(21,3)(21.,13.2)(18.47,15.73)
\qbezier(18.47,15.73)(13.37,20.83)(8.27,15.73)
\qbezier(21,3)(21.,-7.2)(18.47,-9.73)
\qbezier(18.47,-9.73)(13.37,-14.83)(8.27,-9.73)
\qbezier(21,3)(10.8,3.)(8.27,0.47)
\qbezier(8.27,0.47)(3.17,-4.63)(8.27,-9.73)
\qbezier(21,3)(31.2,3.)(33.73,0.47)
\qbezier(33.73,0.47)(38.83,-4.63)(33.73,-9.73)
\qbezier(21,3)(21.,-7.2)(23.53,-9.73)
\qbezier(23.53,-9.73)(28.63,-14.83)(33.73,-9.73)
\epi
&$ 105 $&$ -\ep_8 $&$\ds\f{1}{16}$&$\ds\f{1}{4096}$
\\\hline\hline
5-1 &
\rule[-14pt]{0pt}{34pt}
\bpi(34,0)
\put(17,3){\circle{24}}
\put(8.5,-5.5){\line(1,0){17}}
\put(8.5,-5.5){\line(0,1){17}}
\put(8.5,11.5){\line(1,0){17}}
\put(25.5,-5.5){\line(0,1){17}}
\put(8.5,-5.5){\circle*{2}}
\put(8.5,11.5){\circle*{2}}
\put(25.5,-5.5){\circle*{2}}
\put(25.5,11.5){\circle*{2}}
\epi
&$ 2592 $&$ \ep_4^4 $&$\ds\f{5}{4096}$&$ 7.55133\times10^{-7} $
\\
5-2 &
\rule[-22pt]{0pt}{50pt}
\bpi(42,0)
\put(21,-9){\circle{16}}
\put(21,15){\circle{16}}
\put(13,-9){\line(1,0){16}}
\put(13,15){\line(1,0){16}}
\put(13,3){\oval(16,24)[l]}
\put(29,3){\oval(16,24)[r]}
\put(13,-9){\circle*{2}}
\put(13,15){\circle*{2}}
\put(29,-9){\circle*{2}}
\put(29,15){\circle*{2}}
\epi
&$ 2304 $&$ \ep_4^4 $&$\ds\f{19}{12288}$&$ 1.04187\times10^{-6} $
\\
5-3 &
\rule[-18pt]{0pt}{42pt}
\bpi(58,0)
\put(13,3){\circle{16}}
\put(45,3){\circle{16}}
\put(5,-5){\line(0,1){16}}
\put(25,-5){\oval(40,16)[b]}
\put(25,11){\oval(40,16)[t]}
\put(45,3){\oval(48,16)[l]}
\put(5,3){\circle*{2}}
\put(21,3){\circle*{2}}
\put(45,-5){\circle*{2}}
\put(45,11){\circle*{2}}
\epi
&$ 10368 $&$ \ep_4^4 $&$\ds\f{7}{6144}$&$ 6.71540\times10^{-7} $
\\\hline
5-4 &
\rule[-18pt]{0pt}{42pt}
\bpi(42,0)
\put(21,3){\circle{32}}
\put(5,-13){\oval(32,32)[rt]}
\put(37,-13){\oval(32,32)[lt]}
\put(21,-13){\line(-1,1){16}}
\put(21,-13){\line(1,1){16}}
\put(5,3){\circle*{2}}
\put(37,3){\circle*{2}}
\put(21,-13){\circle*{2}}
\epi
&$ 5760 $&$ -\ep_4^2\ep_6 $&$\ds\f{7}{3072}$&$ 1.50770\times10^{-6} $
\\
5-5 &
\rule[-14pt]{0pt}{50pt}
\bpi(34,0)
\put(17,3){\circle{24}}
\put(17,23){\circle{16}}
\put(6.6,-3){\line(1,0){20.8}}
\put(6.6,-3){\line(3,5){10.4}}
\put(27.4,-3){\line(-3,5){10.4}}
\put(6.6,-3){\circle*{2}}
\put(27.4,-3){\circle*{2}}
\put(17,15){\circle*{2}}
\epi
&$ 12960 $&$ -\ep_4^2\ep_6 $&$\ds\f{3}{1024}$&$ 2.04047\times10^{-6} $
\\
5-6 &
\rule[-14pt]{0pt}{46pt}
\bpi(42,0)
\put(21,3){\circle{24}}
\put(21,3){\oval(24,8)}
\put(9,3){\circle*{2}}
\put(33,3){\circle*{2}}
\put(21,15){\circle*{2}}
\qbezier(21,15)(21.,25.2)(23.53,27.73)
\qbezier(23.53,27.73)(28.63,32.83)(33.73,27.73)
\qbezier(21,15)(31.2,15.)(33.73,17.53)
\qbezier(33.73,17.53)(38.83,22.63)(33.73,27.73)
\qbezier(21,15)(10.8,15.)(8.27,17.53)
\qbezier(8.27,17.53)(3.17,22.63)(8.27,27.73)
\qbezier(21,15)(21.,25.2)(18.47,27.73)
\qbezier(18.47,27.73)(13.37,32.83)(8.27,27.73)
\epi
&$ 4320 $&$ -\ep_4^2\ep_6 $&$\ds\f{5}{1024}$&$ 3.95093\times10^{-6} $
\\\hline
5-7 &
\rule[-22pt]{0pt}{50pt}
\bpi(50,0)
\put(25,3){\circle{40}}
\put(25,3){\oval(40,24)}
\put(25,3){\oval(40,8)}
\put(5,3){\circle*{2}}
\put(45,3){\circle*{2}}
\epi
&$ 360 $&$ \ep_6^2 $&$\ds\f{1}{192}$&$ 3.76084\times10^{-6} $
\\
5-8 &
\rule[-14pt]{0pt}{34pt}
\bpi(66,0)
\put(13,3){\circle{16}}
\put(33,3){\circle{24}}
\put(33,3){\oval(24,8)}
\put(21,3){\circle*{2}}
\put(45,3){\circle*{2}}
\put(53,3){\circle{16}}
\epi
&$ 2700 $&$ \ep_6^2 $&$\ds\f{1}{128}$&$ 6.32149\times10^{-6} $
\\
5-9 &
\rule[-17pt]{0pt}{39pt}
\bpi(58,0)
\put(29,3){\circle{16}}
\put(21,3){\circle*{2}}
\put(37,3){\circle*{2}}
\qbezier(21,3)(10.8,3.)(8.27,5.53)
\qbezier(8.27,5.53)(3.17,10.63)(8.27,15.73)
\qbezier(21,3)(21.,13.2)(18.47,15.73)
\qbezier(18.47,15.73)(13.37,20.83)(8.27,15.73)
\qbezier(21,3)(21.,-7.2)(18.47,-9.73)
\qbezier(18.47,-9.73)(13.37,-14.83)(8.27,-9.73)
\qbezier(21,3)(10.8,3.)(8.27,0.47)
\qbezier(8.27,0.47)(3.17,-4.63)(8.27,-9.73)
\qbezier(37,3)(37.,13.2)(39.53,15.73)
\qbezier(39.53,15.73)(44.63,20.83)(49.73,15.73)
\qbezier(37,3)(47.2,3.)(49.73,5.53)
\qbezier(49.73,5.53)(54.83,10.63)(49.73,15.73)
\qbezier(37,3)(47.2,3.)(49.73,0.47)
\qbezier(49.73,0.47)(54.83,-4.63)(49.73,-9.73)
\qbezier(37,3)(37.,-7.2)(39.53,-9.73)
\qbezier(39.53,-9.73)(44.63,-14.83)(49.73,-9.73)
\epi
&$ 2025 $&$ \ep_6^2 $&$\ds\f{1}{64}$&$\ds\f{1}{65536}$
\\\hline
5-10 &
\rule[-18pt]{0pt}{40pt}
\bpi(50,0)
\put(17,3){\circle{24}}
\put(17,3){\oval(24,8)}
\put(5,3){\circle*{2}}
\put(29,3){\circle*{2}}
\qbezier(29,3)(29.,13.2)(31.53,15.73)
\qbezier(31.53,15.73)(36.63,20.83)(41.73,15.73)
\qbezier(29,3)(39.2,3.)(41.73,5.53)
\qbezier(41.73,5.53)(46.83,10.63)(41.73,15.73)
\qbezier(29,3)(39.2,3.)(41.73,0.47)
\qbezier(41.73,0.47)(46.83,-4.63)(41.73,-9.73)
\qbezier(29,3)(29.,-7.2)(31.53,-9.73)
\qbezier(31.53,-9.73)(36.63,-14.83)(41.73,-9.73)
\epi
&$ 5040 $&$ \ep_4\ep_8 $&$\ds\f{1}{128}$&$ 6.32149\times10^{-6} $
\\\hline
5-11 &
\rule[-17pt]{0pt}{40pt}
\bpi(42,0)
\put(21,3){\circle*{2}}
\qbezier(21,3)(15.6,10.43)(15.6,13.6)
\qbezier(15.6,13.6)(15.6,19.)(21.,19.)
\qbezier(21,3)(26.4,10.43)(26.4,13.6)
\qbezier(26.4,13.6)(26.4,19.)(21.,19.)
\qbezier(21,3)(12.26,0.16)(9.25,1.14)
\qbezier(9.25,1.14)(4.11,2.81)(5.78,7.94)
\qbezier(21,3)(15.6,10.43)(12.59,11.41)
\qbezier(12.59,11.41)(7.45,13.08)(5.78,7.94)
\qbezier(21,3)(21.,-6.19)(19.14,-8.75)
\qbezier(19.14,-8.75)(15.96,-13.12)(11.6,-9.94)
\qbezier(21,3)(12.26,0.16)(10.4,-2.4)
\qbezier(10.4,-2.4)(7.23,-6.77)(11.6,-9.94)
\qbezier(21,3)(29.74,0.16)(31.6,-2.4)
\qbezier(31.6,-2.4)(34.77,-6.77)(30.4,-9.94)
\qbezier(21,3)(21.,-6.19)(22.86,-8.75)
\qbezier(22.86,-8.75)(26.04,-13.12)(30.4,-9.94)
\qbezier(21,3)(26.4,10.43)(29.41,11.41)
\qbezier(29.41,11.41)(34.55,13.08)(36.22,7.94)
\qbezier(21,3)(29.74,0.16)(32.75,1.14)
\qbezier(32.75,1.14)(37.89,2.81)(36.22,7.94)
\epi
&$ 945 $&$ -\ep_{10} $&$\ds\f{1}{32}$&$\ds\f{1}{32768}$
\\\hline\hline
6-1 &
\rule[-20pt]{0pt}{46pt}
\bpi(46,0)
\put(23,3){\circle{36}}
\put(33.58,-11.56){\circle*{2}}
\put(40.12,8.56){\circle*{2}}
\put(23,21){\circle*{2}}
\put(5.88,8.56){\circle*{2}}
\put(12.41,-11.56){\circle*{2}}
\put(12.41,-11.56){\line(1,0){21.17}}
\qbezier(33.58,-11.56)(36.85,-1.5)(40.12,8.56)
\qbezier(40.12,8.56)(31.56,14.78)(23,21)
\qbezier(23,21)(14.44,14.78)(5.88,8.56)
\qbezier(5.88,8.56)(9.15,1.5)(12.41,-11.56)
\epi
&$\ds\f{124416}{5}$&$ -\ep_4^5 $&$\ds\f{35}{131072}$&$ 3.74650 \times 10^{-8} $
\\
6-2 &
\rule[-18pt]{0pt}{42pt}
% used to be \bpi(74,0)
\bpi(66,0)(4,0)
\put(13,3){\circle{16}}
\put(29,3){\circle{16}}
\put(61,3){\circle{16}}
\put(5,-5){\line(0,1){16}}
\put(33,-5){\oval(56,16)[b]}
\put(33,11){\oval(56,16)[t]}
\put(61,3){\oval(48,16)[l]}
\put(5,3){\circle*{2}}
\put(21,3){\circle*{2}}
\put(37,3){\circle*{2}}
\put(61,-5){\circle*{2}}
\put(61,11){\circle*{2}}
\epi
&$ 248832 $&$ -\ep_4^5 $&$\ds\f{71}{294912}$&$ 3.12644 \times 10^{-8} $
\\
6-3 &
\rule[-14pt]{0pt}{46pt}
\bpi(34,0)
\put(5,3){\line(1,0){24}}
\put(17,-9){\line(-1,1){12}}
\put(17,-9){\line(1,1){12}}
\put(17,3){\oval(24,24)[b]}
\put(17,19){\circle{16}}
\put(9,19){\line(1,0){16}}
\put(9,3){\oval(8,32)[tl]}
\put(25,3){\oval(8,32)[tr]}
\put(9,19){\circle*{2}}
\put(25,19){\circle*{2}}
\put(5,3){\circle*{2}}
\put(29,3){\circle*{2}}
\put(17,-9){\circle*{2}}
\epi
&$ 165888 $&$ -\ep_4^5 $&$\ds\f{367}{1179648}$&$ 4.52023 \times 10^{-8} $
\\
6-4 &
\rule[-14pt]{0pt}{34pt}
\bpi(34,0)
\put(17,3){\circle{24}}
\put(8.5,-5.5){\line(1,1){17}}
\put(8.5,-5.5){\line(0,1){17}}
\put(8.5,11.5){\line(1,-1){17}}
\put(25.5,-5.5){\line(0,1){17}}
\put(17,3){\circle*{2}}
\put(8.5,-5.5){\circle*{2}}
\put(8.5,11.5){\circle*{2}}
\put(25.5,-5.5){\circle*{2}}
\put(25.5,11.5){\circle*{2}}
\epi
&$ 497664 $&$ -\ep_4^5 $&$\ds\f{269}{1179648}$&$ 2.85447 \times 10^{-8} $
\\
6-5 &
\rule[-20pt]{0pt}{46pt}
\bpi(46,0)
\put(23,3){\circle{36}}
\put(33.58,-11.56){\circle*{2}}
\put(40.12,8.56){\circle*{2}}
\put(23,21){\circle*{2}}
\put(5.88,8.56){\circle*{2}}
\put(12.41,-11.56){\circle*{2}}
\put(5.88,8.56){\line(1,0){34.24}}
\qbezier(33.58,-11.56)(19.73,-1.5)(5.88,8.56)
\qbezier(33.58,-11.56)(28.29, 4.72)(23,21)
\qbezier(12.41,-11.56)(26.27,-1.5)(40.12,8.56)
\qbezier(12.41,-11.56)(17.71,4.72)(23,21)
\epi
&$\ds\f{331776}{5}$&$ -\ep_4^5 $&$\ds\f{5}{24576}$
&$2.37861(5)\!\times\!10^{-8}$
\\
6-6 &
\rule[-14pt]{0pt}{34pt}
\bpi(58,0)
\put(17,3){\circle{24}}
\put(17,3){\oval(8,24)}
\put(17,-9){\circle*{2}}
\put(17,15){\circle*{2}}
\put(41,3){\circle{24}}
\put(41,3){\oval(8,24)}
\put(41,-9){\circle*{2}}
\put(41,15){\circle*{2}}
\put(29,3){\circle*{2}}
\epi
&$ 27648 $&$ -\ep_4^5 $&$\ds\f{25}{65536}$&$ 6.39380 \times 10^{-8} $
\\\hline
6-7 &
\rule[-20pt]{0pt}{46pt}
\bpi(46,0)
\put(23,3){\circle{36}}
\put(5,3){\circle*{2}}
\put(41,3){\circle*{2}}
\put(23,-15){\circle*{2}}
\put(23,21){\circle*{2}}
\put(5,3){\line(1,0){36}}
\put(23,-15){\line(-1,1){16}}
\put(23,21){\line(-1,-1){16}}
\put(23,21){\line(1,-1){16}}
\put(5,-15){\oval(36,36)[tr]}
\epi
&$ 414720 $&$ \ep_4^3\ep_6 $&$\ds\f{65}{147456}$&$ 6.27924\times 10^{-8} $
\\
6-8 &
\rule[-20pt]{0pt}{46pt}
\bpi(46,0)
\put(23,3){\circle*{2}}
\put(23,21){\circle*{2}}
\put(7.41,-6.){\circle*{2}}
\put(38.59,-6.){\circle*{2}}
\put(23,3){\circle{36}}
\qbezier(23,3)(15.79,10.21)(15.79,13.79)
\qbezier(15.79,13.79)(15.79,21.)(23.,21.)
\qbezier(23,3)(30.21,10.21)(30.21,13.79)
\qbezier(30.21,13.79)(30.21,21.)(23.,21.)
\qbezier(23,3)(20.36,-6.85)(17.26,-8.64)
\qbezier(17.26,-8.64)(11.02,-12.24)(7.41,-6.)
\qbezier(23,3)(13.15,5.64)(10.05,3.85)
\qbezier(10.05,3.85)(3.81,0.24)(7.41,-6.)
\qbezier(23,3)(32.85,5.64)(35.95,3.85)
\qbezier(35.95,3.85)(42.19,0.24)(38.59,-6.)
\qbezier(23,3)(25.64,-6.85)(28.74,-8.64)
\qbezier(28.74,-8.64)(34.98,-12.24)(38.59,-6.)
\epi
&$ 207360 $&$ \ep_4^3\ep_6 $&$\ds\f{5}{12288}$&$ 5.50218 \times 10^{-8} $
\\
6-9 &
\rule[-22pt]{0pt}{50pt}
\bpi(42,0)
\put(21,-9){\circle{16}}
\put(21,15){\circle{16}}
\put(13,-9){\line(1,0){16}}
\put(13,15){\line(1,0){16}}
\put(13,3){\oval(16,24)[l]}
\put(29,3){\oval(16,24)[r]}
\put(13,-9){\circle*{2}}
\put(13,15){\circle*{2}}
\put(29,-9){\circle*{2}}
\put(29,15){\circle*{2}}
\qbezier(29,15)(29.,21.8)(30.69,23.49)
\qbezier(30.69,23.49)(34.09,26.88)(37.49,23.49)
\qbezier(29,15)(35.8,15.)(37.49,16.69)
\qbezier(37.49,16.69)(40.88,20.09)(37.49,23.49)
\epi
&$ 138240 $&$ \ep_4^3\ep_6 $&$\ds\f{19}{24576}$&$ 1.30234 \times 10^{-7} $
\\
6-10 &
\rule[-14pt]{0pt}{34pt}
\bpi(50,0)
\put(17,3){\circle{24}}
\put(5,3){\line(1,1){12}}
\put(5,3){\line(1,-1){12}}
\put(29,3){\line(-1,1){12}}
\put(29,3){\line(-1,-1){12}}
\put(5,3){\circle*{2}}
\put(29,3){\circle*{2}}
\put(17,-9){\circle*{2}}
\put(17,15){\circle*{2}}
\put(37,3){\circle{16}}
\epi
&$ 155520 $&$ \ep_4^3\ep_6 $&$\ds\f{5}{8192}$&$ 9.43917 \times 10^{-8} $
\\
6-11 &
\rule[-18pt]{0pt}{42pt}
% used to be \bpi(74,0)
\bpi(66,0)(4,0)
\put(13,3){\circle{16}}
\put(29,3){\circle{16}}
\put(61,3){\circle{16}}
\put(21,-5){\line(0,1){16}}
\put(41,-5){\oval(40,16)[b]}
\put(41,11){\oval(40,16)[t]}
\put(61,3){\oval(48,16)[l]}
\put(21,3){\circle*{2}}
\put(37,3){\circle*{2}}
\put(61,-5){\circle*{2}}
\put(61,11){\circle*{2}}
\epi
&$ 622080 $&$ \ep_4^3\ep_6 $&$\ds\f{7}{12288}$&$ 8.39425 \times 10^{-8}  $
\\
6-12 &
\rule[-14pt]{0pt}{46pt}
\bpi(42,0)
\put(21,3){\circle{24}}
\qbezier(21,15)(21.,25.2)(23.53,27.73)
\qbezier(23.53,27.73)(28.63,32.83)(33.73,27.73)
\qbezier(21,15)(31.2,15.)(33.73,17.53)
\qbezier(33.73,17.53)(38.83,22.63)(33.73,27.73)
\qbezier(21,15)(10.8,15.)(8.27,17.53)
\qbezier(8.27,17.53)(3.17,22.63)(8.27,27.73)
\qbezier(21,15)(21.,25.2)(18.47,27.73)
\qbezier(18.47,27.73)(13.37,32.83)(8.27,27.73)
\put(10.6,9){\line(1,0){20.8}}
\put(10.6,9){\line(3,-5){10.4}}
\put(31.4,9){\line(-3,-5){10.4}}
\put(10.6,9){\circle*{2}}
\put(31.4,9){\circle*{2}}
\put(21,-9){\circle*{2}}
\put(21,15){\circle*{2}}
\epi
&$ 155520 $&$ \ep_4^3\ep_6 $&$\ds\f{1}{1024}$&$ 1.70039 \times 10^{-7} $
\\
6-13 &
\rule[-14pt]{0pt}{34pt}
\bpi(58,0)
\put(17,3){\circle{24}}
\put(17,3){\oval(24,8)}
\put(5,3){\circle*{2}}
\put(41,3){\circle{24}}
\put(41,3){\oval(8,24)}
\put(41,-9){\circle*{2}}
\put(41,15){\circle*{2}}
\put(29,3){\circle*{2}}
\epi
&$ 34560 $&$ \ep_4^3\ep_6 $&$\ds\f{5}{8192}$&$ 1.02301 \times 10^{-7} $
\\
6-14 &
\rule[-18pt]{0pt}{40pt}
\bpi(66,0)
\put(17,3){\circle{24}}
\put(17,3){\oval(8,24)}
\put(17,-9){\circle*{2}}
\put(17,15){\circle*{2}}
\put(29,3){\circle*{2}}
\put(37,3){\circle{16}}
\put(45,3){\circle*{2}}
\qbezier(45,3)(45.,13.2)(47.53,15.73)
\qbezier(47.53,15.73)(52.63,20.83)(57.73,15.73)
\qbezier(45,3)(55.2,3.)(57.73,5.53)
\qbezier(57.73,5.53)(62.83,10.63)(57.73,15.73)
\qbezier(45,3)(55.2,3.)(57.73,0.47)
\qbezier(57.73,0.47)(62.83,-4.63)(57.73,-9.73)
\qbezier(45,3)(45.,-7.2)(47.53,-9.73)
\qbezier(47.53,-9.73)(52.63,-14.83)(57.73,-9.73)
\epi
&$ 51840 $&$ \ep_4^3\ep_6 $&$\ds\f{5}{4096}$&$ 2.46933 \times 10^{-7} $
\\\hline
6-15 &
\rule[-18pt]{0pt}{42pt}
\bpi(58,0)
\put(21,3){\circle{32}}
\put(37,19){\oval(32,32)[lb]}
\put(37,-13){\oval(32,32)[lt]}
\put(21,19){\line(1,-1){16}}
\put(21,-13){\line(1,1){16}}
\put(21,19){\circle*{2}}
\put(37,3){\circle*{2}}
\put(21,-13){\circle*{2}}
\put(45,3){\circle{16}}
\epi
&$ 161280 $&$ -\ep_4^2\ep_8 $&$\ds\f{7}{6144}$&$ 1.88463 \times 10^{-7} $
\\
6-16 &
\rule[-14pt]{0pt}{46pt}
\bpi(42,0)
\put(21,3){\circle{24}}
\qbezier(21,15)(21.,25.2)(23.53,27.73)
\qbezier(23.53,27.73)(28.63,32.83)(33.73,27.73)
\qbezier(21,15)(31.2,15.)(33.73,17.53)
\qbezier(33.73,17.53)(38.83,22.63)(33.73,27.73)
\qbezier(21,15)(10.8,15.)(8.27,17.53)
\qbezier(8.27,17.53)(3.17,22.63)(8.27,27.73)
\qbezier(21,15)(21.,25.2)(18.47,27.73)
\qbezier(18.47,27.73)(13.37,32.83)(8.27,27.73)
\put(10.6,-3){\line(1,0){20.8}}
\put(10.6,-3){\line(3,5){10.4}}
\put(31.4,-3){\line(-3,5){10.4}}
\put(10.6,-3){\circle*{2}}
\put(31.4,-3){\circle*{2}}
\put(21,15){\circle*{2}}
\epi
&$ 181440 $&$ -\ep_4^2\ep_8 $&$\ds\f{3}{2048}$&$ 2.55058 \times 10^{-7} $
\\
6-17 &
\rule[-14pt]{0pt}{34pt}
\bpi(58,0)
\put(17,3){\circle{24}}
\put(17,3){\oval(24,8)}
\put(41,3){\circle{24}}
\put(41,3){\oval(24,8)}
\put(5,3){\circle*{2}}
\put(29,3){\circle*{2}}
\put(53,3){\circle*{2}}
\epi
&$ 20160 $&$ -\ep_4^2\ep_8 $&$\ds\f{1}{1024}$&$ 1.63681 \times 10^{-7} $
\\
6-18 &
\rule[-14pt]{0pt}{52pt}
\bpi(42,0)
\put(21,3){\circle{24}}
\put(21,3){\oval(24,8)}
\put(9,3){\circle*{2}}
\put(33,3){\circle*{2}}
\put(21,15){\circle*{2}}
\qbezier(21,15)(26.27,24.13)(29.39,25.93)
\qbezier(29.39,25.93)(33.95,28.57)(36.59,24.)
\qbezier(21,15)(31.54,15.)(34.66,16.8)
\qbezier(34.66,16.8)(39.22,19.43)(36.59,24.)
\qbezier(21,15)(15.73,24.13)(15.73,27.73)
\qbezier(15.73,27.73)(15.73,33.)(21.,33.)
\qbezier(21,15)(26.27,24.13)(26.27,27.73)
\qbezier(26.27,27.73)(26.27,33.)(21.,33.)
\qbezier(21,15)(10.46,15.)(7.34,16.8)
\qbezier(7.34,16.8)(2.78,19.43)(5.41,24.)
\qbezier(21,15)(15.73,24.13)(12.61,25.93)
\qbezier(12.61,25.93)(8.05,28.57)(5.41,24.)
\epi
&$ 40320 $&$ -\ep_4^2\ep_8 $&$\ds\f{5}{2048}$&$ 4.93867 \times 10^{-7} $
\\\hline
6-19 &
\rule[-22pt]{0pt}{50pt}
\bpi(50,0)
\put(25,3){\circle{40}}
\put(25,3){\oval(40,24)}
\put(15,3){\oval(20,8)}
\put(35,3){\oval(20,8)}
\put(5,3){\circle*{2}}
\put(25,3){\circle*{2}}
\put(45,3){\circle*{2}}
\epi
&$ 64800 $&$ -\ep_4\ep_6^2 $&$\ds\f{1}{1152}$&$ 1.31735 \times 10^{-7} $
\\
6-20 &
\rule[-18pt]{0pt}{42pt}
\bpi(58,0)
\put(21,3){\circle{32}}
\put(5,-13){\oval(32,32)[rt]}
\put(37,-13){\oval(32,32)[lt]}
\put(21,-13){\line(-1,1){16}}
\put(21,-13){\line(1,1){16}}
\put(5,3){\circle*{2}}
\put(37,3){\circle*{2}}
\put(21,-13){\circle*{2}}
\put(45,3){\circle{16}}
\epi
&$ 172800 $&$ -\ep_4\ep_6^2 $&$\ds\f{7}{6144}$&$ 1.88463 \times 10^{-7} $
\\
6-21 &
\rule[-14pt]{0pt}{40pt}
\bpi(60.66,0)(-13.33,0)
\put(17,3){\circle{24}}
\put(6.6,9){\line(1,0){20.8}}
\put(6.6,9){\line(3,-5){10.4}}
\put(27.4,9){\line(-3,-5){10.4}}
\put(6.6,9){\circle*{2}}
\put(27.4,9){\circle*{2}}
\put(17,-9){\circle*{2}}
\put(-0.33,13){\circle{16}}
\put(34.33,13){\circle{16}}
\epi
&$ 194400 $&$ -\ep_4\ep_6^2 $&$\ds\f{3}{2048}$&$ 2.55058 \times 10^{-7} $
\\
6-22 &
\rule[-18pt]{0pt}{40pt}
\bpi(66,0)
\put(17,3){\circle{24}}
\put(17,3){\oval(24,8)}
\put(5,3){\circle*{2}}
\put(29,3){\circle*{2}}
\put(37,3){\circle{16}}
\put(45,3){\circle*{2}}
\qbezier(45,3)(45.,13.2)(47.53,15.73)
\qbezier(47.53,15.73)(52.63,20.83)(57.73,15.73)
\qbezier(45,3)(55.2,3.)(57.73,5.53)
\qbezier(57.73,5.53)(62.83,10.63)(57.73,15.73)
\qbezier(45,3)(55.2,3.)(57.73,0.47)
\qbezier(57.73,0.47)(62.83,-4.63)(57.73,-9.73)
\qbezier(45,3)(45.,-7.2)(47.53,-9.73)
\qbezier(47.53,-9.73)(52.63,-14.83)(57.73,-9.73)
\epi
&$ 32400 $&$ -\ep_4\ep_6^2 $&$\ds\f{1}{512}$&$ 3.95093 \times 10^{-7}  $
\\
6-23 &
\rule[-14pt]{0pt}{46pt}
\bpi(59,0)
\put(21,3){\circle{24}}
\put(21,3){\oval(24,8)}
\put(9,3){\circle*{2}}
\put(33,3){\circle*{2}}
\put(21,15){\circle*{2}}
\put(41,3){\circle{16}}
\qbezier(21,15)(21.,25.2)(23.53,27.73)
\qbezier(23.53,27.73)(28.63,32.83)(33.73,27.73)
\qbezier(21,15)(31.2,15.)(33.73,17.53)
\qbezier(33.73,17.53)(38.83,22.63)(33.73,27.73)
\qbezier(21,15)(10.8,15.)(8.27,17.53)
\qbezier(8.27,17.53)(3.17,22.63)(8.27,27.73)
\qbezier(21,15)(21.,25.2)(18.47,27.73)
\qbezier(18.47,27.73)(13.37,32.83)(8.27,27.73)
\epi
&$ 129600 $&$ -\ep_4\ep_6^2 $&$\ds\f{5}{2048}$&$ 4.93867 \times 10^{-7} $
\\
6-24 &
\rule[-17pt]{0pt}{39pt}
% used to be \bpi(74,0)
\bpi(66,0)(4,0)
\put(29,3){\circle{16}}
\put(45,3){\circle{16}}
\put(21,3){\circle*{2}}
\put(37,3){\circle*{2}}
\put(53,3){\circle*{2}}
\qbezier(21,3)(10.8,3.)(8.27,5.53)
\qbezier(8.27,5.53)(3.17,10.63)(8.27,15.73)
\qbezier(21,3)(21.,13.2)(18.47,15.73)
\qbezier(18.47,15.73)(13.37,20.83)(8.27,15.73)
\qbezier(21,3)(21.,-7.2)(18.47,-9.73)
\qbezier(18.47,-9.73)(13.37,-14.83)(8.27,-9.73)
\qbezier(21,3)(10.8,3.)(8.27,0.47)
\qbezier(8.27,0.47)(3.17,-4.63)(8.27,-9.73)
\qbezier(53,3)(53.,13.2)(55.53,15.73)
\qbezier(55.53,15.73)(60.63,20.83)(65.73,15.73)
\qbezier(53,3)(63.2,3.)(65.73,5.53)
\qbezier(65.73,5.53)(70.83,10.63)(65.73,15.73)
\qbezier(53,3)(63.2,3.)(65.73,0.47)
\qbezier(65.73,0.47)(70.83,-4.63)(65.73,-9.73)
\qbezier(53,3)(53.,-7.2)(55.53,-9.73)
\qbezier(55.53,-9.73)(60.63,-14.83)(65.73,-9.73)
\epi
&$ 24300 $&$ -\ep_4\ep_6^2 $&$\ds\f{1}{256}$&$\ds\f{1}{1048576}$
\\\hline
6-25 &
\rule[-19pt]{0pt}{44pt}
\bpi(52,0)
\put(17,3){\circle{24}}
\put(17,3){\oval(24,8)}
\put(5,3){\circle*{2}}
\put(29,3){\circle*{2}}
\qbezier(29,3)(38.13,-2.27)(39.93,-5.39)
\qbezier(39.93,-5.39)(42.57,-9.95)(38.,-12.59)
\qbezier(29,3)(29.,-7.54)(30.8,-10.66)
\qbezier(30.8,-10.66)(33.43,-15.22)(38.,-12.59)
\qbezier(29,3)(38.13,8.27)(41.73,8.27)
\qbezier(41.73,8.27)(47.,8.27)(47.,3.)
\qbezier(29,3)(38.13,-2.27)(41.73,-2.27)
\qbezier(41.73,-2.27)(47.,-2.27)(47.,3.)
\qbezier(29,3)(29.,13.54)(30.8,16.66)
\qbezier(30.8,16.66)(33.43,21.22)(38.,18.59)
\qbezier(29,3)(38.13,8.27)(39.93,11.39)
\qbezier(39.93,11.39)(42.57,15.95)(38.,18.59)
\epi
&$ 75600 $&$ \ep_4\ep_{10} $&$\ds\f{1}{256}$&$ 7.90187 \times 10^{-7}  $
\\\hline
6-26 &
\rule[-22pt]{0pt}{50pt}
\bpi(66,0)
\put(25,3){\circle{40}}
\put(25,3){\oval(40,24)}
\put(25,3){\oval(40,8)}
\put(5,3){\circle*{2}}
\put(45,3){\circle*{2}}
\put(53,3){\circle{16}}
\epi
&$ 20160 $&$ \ep_6\ep_8 $&$\ds\f{1}{384}$&$ 4.70105 \times 10^{-7} $
\\
6-27 &
\rule[-18pt]{0pt}{42pt}
\bpi(66,0)
\put(13,3){\circle{16}}
\put(33,3){\circle{24}}
\put(33,3){\oval(24,8)}
\put(21,3){\circle*{2}}
\put(45,3){\circle*{2}}
\qbezier(45,3)(55.2,3.)(57.73,0.47)
\qbezier(57.73,0.47)(62.83,-4.63)(57.73,-9.73)
\qbezier(45,3)(45.,-7.2)(47.53,-9.73)
\qbezier(47.53,-9.73)(52.63,-14.83)(57.73,-9.73)
\qbezier(45,3)(45.,13.2)(47.53,15.73)
\qbezier(47.53,15.73)(52.63,20.83)(57.73,15.73)
\qbezier(45,3)(55.2,3.)(57.73,5.53)
\qbezier(57.73,5.53)(62.83,10.63)(57.73,15.73)
\epi
&$ 75600 $&$ \ep_6\ep_8 $&$\ds\f{1}{256}$&$ 7.90187 \times 10^{-7} $
\\
6-28 &
\rule[-17pt]{0pt}{39pt}
\bpi(60,0)
\put(29,3){\circle{16}}
\put(21,3){\circle*{2}}
\put(37,3){\circle*{2}}
\qbezier(21,3)(10.8,3.)(8.27,5.53)
\qbezier(8.27,5.53)(3.17,10.63)(8.27,15.73)
\qbezier(21,3)(21.,13.2)(18.47,15.73)
\qbezier(18.47,15.73)(13.37,20.83)(8.27,15.73)
\qbezier(21,3)(21.,-7.2)(18.47,-9.73)
\qbezier(18.47,-9.73)(13.37,-14.83)(8.27,-9.73)
\qbezier(21,3)(10.8,3.)(8.27,0.47)
\qbezier(8.27,0.47)(3.17,-4.63)(8.27,-9.73)
\qbezier(37,3)(37.,13.54)(38.8,16.66)
\qbezier(38.8,16.66)(41.43,21.22)(46.,18.59)
\qbezier(37,3)(46.13,8.27)(47.93,11.39)
\qbezier(47.93,11.39)(50.57,15.95)(46.,18.59)
\qbezier(37,3)(46.13,8.27)(49.73,8.27)
\qbezier(49.73,8.27)(55.,8.27)(55.,3.)
\qbezier(37,3)(46.13,-2.27)(49.73,-2.27)
\qbezier(49.73,-2.27)(55.,-2.27)(55.,3.)
\qbezier(37,3)(46.13,-2.27)(47.93,-5.39)
\qbezier(47.93,-5.39)(50.57,-9.95)(46.,-12.59)
\qbezier(37,3)(37.,-7.54)(38.8,-10.66)
\qbezier(38.8,-10.66)(41.43,-15.22)(46.,-12.59)
\epi
&$ 37800 $&$ \ep_6\ep_8 $&$\ds\f{1}{128}$&$\ds\f{1}{524288}$
\\\hline
6-29 &
\rule[-18pt]{0pt}{42pt}
\bpi(42,0)
\put(21,3){\circle*{2}}
\qbezier(21,3)(25.69,11.12)(28.45,12.72)
\qbezier(28.45,12.72)(32.51,15.06)(34.86,11.)
\qbezier(21,3)(30.37,3.)(33.14,4.6)
\qbezier(33.14,4.6)(37.2,6.94)(34.86,11.)
\qbezier(21,3)(16.31,11.12)(16.31,14.31)
\qbezier(16.31,14.31)(16.31,19.)(21.,19.)
\qbezier(21,3)(25.69,11.12)(25.69,14.31)
\qbezier(25.69,14.31)(25.69,19.)(21.,19.)
\qbezier(21,3)(11.63,3.)(8.86,4.6)
\qbezier(8.86,4.6)(4.8,6.94)(7.14,11.)
\qbezier(21,3)(16.31,11.12)(13.55,12.72)
\qbezier(13.55,12.72)(9.49,15.06)(7.14,11.)
\qbezier(21,3)(16.31,-5.12)(13.55,-6.72)
\qbezier(13.55,-6.72)(9.49,-9.06)(7.14,-5.)
\qbezier(21,3)(11.63,3.)(8.86,1.4)
\qbezier(8.86,1.4)(4.8,-0.94)(7.14,-5.)
\qbezier(21,3)(25.69,-5.12)(25.69,-8.31)
\qbezier(25.69,-8.31)(25.69,-13.)(21.,-13.)
\qbezier(21,3)(16.31,-5.12)(16.31,-8.31)
\qbezier(16.31,-8.31)(16.31,-13.)(21.,-13.)
\qbezier(21,3)(30.37,3.)(33.14,1.4)
\qbezier(33.14,1.4)(37.2,-0.94)(34.86,-5.)
\qbezier(21,3)(25.69,-5.12)(28.45,-6.72)
\qbezier(28.45,-6.72)(32.51,-9.06)(34.86,-5.)
\epi
&$ 10395 $&$ -\ep_{12} $&$\ds\f{1}{64}$&$\ds\f{1}{262144}$
\\\hline\hline
\end{longtable}
\endgroup

\end{document}